\documentclass[10pt]{article}
\usepackage{amssymb, amsmath, amsthm, bm, float, graphicx,  tabularx, geometry, mathrsfs, setspace, enumerate, appendix, csquotes, caption, url, yfonts, pslatex, sectsty, eulervm, xcolor, hyperref}
\usepackage[T1]{fontenc}
\usepackage[round]{natbib}
\usepackage{pgf}
\usepackage{tikz}
\usetikzlibrary{arrows, automata, positioning}
\newtheorem{theorem}{Theorem}

\newtheorem{definition}{Definition}
\newtheorem{example}{Example}

\newtheorem{lemma}{Lemma}

\newtheorem{remark}{Remark}

\newcommand{\R}{\mathbb{R}}
\newcommand{\Q}{\mathbb{Q}}

\newcommand{\N}{\mathbb{N}}

\input{tcilatex}
\begin{document}

\title{On social welfare orders satisfying anonymity and asymptotic density-one Pareto}
\author{Ram Sewak Dubey\thanks{%
Department of Economics, Feliciano School of Business, Montclair State University, Montclair, NJ 07043; E-mail: dubeyr@montclair.edu} \and%
Giorgio Laguzzi\thanks{%
University of Freiburg in the Mathematical Logic Group at Eckerstr. 1, 79104 Freiburg im Breisgau, Germany; Email: giorgio.laguzzi@libero.it} \and %
Francesco Ruscitti\thanks{%
Department of Economics and Social Sciences, John Cabot University, Via della Lungara 233, 00165 Rome, Italy. Email: fruscitti@johncabot.edu}}
\date{\today}
\maketitle

\begin{abstract}

We study the nature (constructive versus non-constructive) and the issue of real-valued representability of social welfare orders, on the set of infinite utility streams, satisfying the anonymity and asymptotic density-one Pareto axioms. 
We characterize the existence of representable and constructive social welfare orders (fulfilling the aforementioned axioms) in terms of easily verifiable conditions on the feasible set of one-period utilities, denoted by $Y\subset \R$: a social welfare order satisfying anonymity and asymptotic density-one Pareto is representable as well as admits explicit description if and only if $Y$ contains finitely many elements.

\noindent \emph{Keywords:} \texttt{Anonymity,}\; \texttt{Asymptotic density-one Pareto,}\; \texttt{Non-Ramsey set,}\;\texttt{Social welfare order.}

\noindent \emph{Journal of Economic Literature} Classification Numbers: 
\texttt{D60,}\; \texttt{D70,}\; \texttt{D90.}
\end{abstract}

\setcounter{MaxMatrixCols}{10}

\newpage

\section{Introduction}
\label{S1}

This paper provides a contribution to the literature on \emph{intergenerational equity}, which has drawn much attention from economists, philosophers and social planners. 
One of the principal issues examined in these studies is how to take the future generations' well-being into account relative to the well-being of the current generations.
A formal discussion of the concept of intergenerational equity has a long history in the economics literature. 
\citet{ramsey1928} observed that discounting one generation's utility relative to another's is \enquote{ethically indefensible}, and something that \enquote{arises merely from the weakness of the imagination}. 
In the sixties, Ramsey's intuitive idea found support among scholars in the field of inter-temporal social choice theory. 
\citet{diamond1965} formalized it as the concept of \enquote{equal treatment} of all generations (present and future) and termed it the \emph{anonymity} axiom on social preferences; this axiom requires the social planner to be indifferent between any pair of utility streams if any one of them is obtained from the other by interchanging the levels of well-being of any two generations.
Yet, as is well known, a vast body of the dynamic programming literature is based on the assumption that the objective function is a discounted sum of one-period return functions.
However, it is important to note that, given any positive discount rate, the method of discounting down-weighs policy consequences that occur in the future, thus undermining the extent to which we care for the welfare of future generations.

Increasingly, economists are therefore being asked to evaluate and formulate policies with a planning horizon that extends over the infinite future. 
There are plenty of prominent examples: global climate change, non-renewable natural resources, radioactive waste disposal, loss of biodiversity, groundwater pollution, minerals depletion, and many others. 
This task entails consistently comparing infinite utility streams by means of social preferences that respect certain desirable axioms. 
Social preferences come in the form of a binary relation (i.e., a reflexive and transitive pair-wise ranking rule) called social welfare relation. 
The latter is referred to as a social welfare order when the binary relation at hand is also complete (i.e., capable of ranking any pair of infinite utility streams). 
Furthermore, a numerical representation of a social welfare order is known as social welfare function. 
The subject of consistent evaluation has received significant attention in the research community. 

The standard framework in the literature treats infinite utility streams as elements of the set $X\equiv Y^{\mathbb{N}}$, where $Y\subset \mathbb{R}$ is a non-empty subset of real numbers and $\mathbb{N}$ is the set of natural numbers. 
The set $Y$ describes all possible levels of utility that any generation can attain. 
Observe that aggregating social preferences that satisfy the anonymity axiom presents no challenges. 
Indeed, a social welfare function which assigns the same value to all utility sequences is trivially anonymous. 
However, such a social welfare function is of no interest to the social planner as it does not convey any meaningful information. 
Clearly, we want to look at social preferences that are able to detect changes in utility profiles. 
To this end, what is needed is some degree of sensitivity (to the levels of individuals' well-being across infinitely-many generations), which is captured by the \emph{Pareto} axiom. 
The latter has achieved general consensus among economists and policy makers. 
This axiom requires a stream of utilities to be ranked above another stream if at least one generation receives a higher utility and no generation receives lower utility compared to the other stream. 
It is a basic efficiency concept which is generally applied in models of economic growth. 
We will refer to social preferences satisfying the anonymity (equity) and Pareto (efficiency) axioms as  \emph{equitable} or \emph{ethical} preferences.

The results of this paper fit well in the body of research studying various versions of the Pareto principles combined with anonymity. 
In short, we introduce a new efficiency criterion, we refer to as asymptotic density-one Pareto, and we examine social welfare relations on $X\equiv Y^{\mathbb{N}}$ which are complete and satisfy the anonymity and asymptotic density-one Pareto conditions. 
We show that the restrictions on $Y$ that lead to the social welfare order at hand having a real-valued representation are the same as the restrictions on $Y$ that result in the social welfare order admitting explicit description. 
More specifically, each of the two properties hold if and only if $Y\subset \mathbb{R}$ is finite. 
Specifically, our results extend those in \citet{petri2019} and \citet{dubey2020}.
We next provide a brief overview of the recent literature on social choice and welfare and discuss more in detail the scope and position of our contribution within the established literature.%
\footnote{See \citet{asheim2010} for an excellent survey of the literature on intergenerational equity.} 

\subsection{Historical background and latest developments} \label{ss11}

In an influential paper, \citet{diamond1965} discovered that a conflict between the anonymity and Pareto axioms arises if one seeks a continuous social welfare order, where continuity is defined with respect to the sup metric. 
He assumed the set $Y$ to be the closed interval $[0,1]$ and proved non-existence of a continuous social welfare order which respects the anonymity and Pareto axioms. 
However, if the set $Y$ is the closed interval $[0,1]$, a real-valued representation of a social welfare order satisfying the Pareto and continuity conditions does exist.%
\footnote{To see this, consider the social welfare function $W(x)=\sum_{t=1}^{\infty }\delta ^{t-1}x_{t}$, with $\delta \in(0,1)$. 
It is Paretian and continuous with respect to the sup-metric. \label{f2}}
Therefore, it can be inferred from Diamond's result and the Paretian social welfare function (see footnote \ref{f2}) that, when $Y$ is the closed interval $[0, 1]$, there does not exist any social welfare function satisfying the anonymity and Pareto axioms which is continuous in the sup-metric. 
\citet{basu2003} refined Diamond's result by showing that the clash between representability and the anonymity and Pareto axioms persists even when the continuity axiom is discarded. 
Their impossibility result established that there does not exist any social welfare function satisfying the anonymity and Pareto axioms if $Y$ contains at least two distinct elements. 
In other words, there does not exist any representable social welfare order satisfying the anonymity and Pareto axioms when $Y$ is non-trivial.%
\footnote{In case $Y$ contains one element, the problem at hand is vacuous  as the set $X$ is a singleton, i.e., the constant sequence.} 
Hence, demanding a real-valued representation forces one to drop the continuity condition (with respect to sup-norm metric) in Diamond's result. 
It is worthwhile to note that in \citet{basu2003} the requirement on the set $Y$ is minimal (i.e., $|Y|=2$).

We learn from \citet{diamond1965} and \citet{basu2003} that the pursuit of equitable social welfare functions requires weakening the Pareto axiom. 
Weakening the Pareto condition to weak Pareto%
\footnote{The weak Pareto axiom requires a stream of utilities to be ranked above another one if every generation is better off in the former compared to the latter.} 
leads to both possibility and impossibility results, depending on the structure of the utility domain $Y$. 
\citet[Theorem 3]{mitra2007c} prove that the function $W(x)=\min \left\{x_{t}\right\}$, defined on the set $X$ where $Y$ is the set of natural numbers, represents a social welfare order satisfying anonymity and weak Pareto. 
On the other hand, \citet[Theorem 4]{mitra2007c} show that there does not exist any social welfare function satisfying anonymity and weak Pareto when $Y$ contains the interval $[0,1]$.

In the subsequent literature, scholars focused their attention on intermediate degrees of sensitivity (in-between the two extremes of Pareto and weak Pareto) in the effort to refine the possibility/impossibility divide regarding the representation of ethical social welfare orders. 
The present work offers a contribution to this branch of the literature.
To explain the significance of our results, a brief overview of the established literature is in order. 
\citet{crespo2009} examine the conflict between representability, anonymity, and a weaker form of the Pareto condition, namely, the infinite Pareto axiom. 
The infinite Pareto axiom states that, for a given utlity stream, an increase in the utilty level of infinitely many generations (all else being equal) leads to a new stream which is socially preferred to the initial one. 
Even though infinite Pareto is considerably weaker than the Pareto condition, the impossibility result obtained in \citet{crespo2009} still holds true for all non-trivial domains $Y$, as was shown by \citet{basu2003}.
\citet{dubey2011} characterize the restrictions on $Y$ for the existence of social welfare functions satisfying the weak Pareto and anonymity axioms.
Their paper reveals that the crucial feature of $Y$ (for the existence of social welfare orders satisfying a set of desiderata) is the order type of its subsets of real numbers, as opposed to the mere cardinality of the set $Y$, a fact which could be intuitively inferred from \citet{mitra2007c}. 
Indeed, a social welfare function satisfying the weak Pareto and anonymity axioms exists if and only if the set $Y$ does not contain a subset of order type similar to the set of integers.

\citet{petri2019} notes that on the one hand there is no anonymous social welfare function satisfying infinite Pareto (established in \citet{crespo2009}) for all non-trivial $Y$, and on the other hand anonymous social welfare functions satisfying weak Pareto exist even when $Y$ contains countably many elements (i.e., the set of natural numbers). 
This suggests, naturally, that one may want to treat the version of the Pareto axiom (to be used) as a variable and investigate how weak the Pareto axiom needs to be in order for an anonymous social welfare order to admit a representation on a non-trivial domain $Y$. 
Along these lines, \citet{petri2019} uncovered a version of the Pareto axiom, which yields existence of an anonymous Paretian social welfare function, based on a new insight. 
He came up with the lower asymptotic Pareto axiom, which relies on the notion of asymptotic density of subsets of the natural numbers. 
The lower asymptotic Pareto axiom requires a stream of utilities to be ranked above another one if under the former generations belonging to a subset with positive lower asymptotic density are better off than under the latter, and no one has lower utility in the former compared to the latter. 
This is a significant result as it helps evaluate infinite utility streams with a real-valued function which respects the anonymity axiom and the newly discovered version of the Pareto axiom for sets $Y$ which could contain a large (but finite) number of elements. 
It is a characterization theorem in the sense that for all $Y$ with infinite cardinality an impossibility result is proven in \citet{petri2019}. 
The axiom of weak Pareto (which hinges on a welfare improvement ocurring for every generation) could in some sense be too demanding, but the lower asymptotic Pareto condition applies even when the welfare of infinitely many generations remain unchanged (across the pair of utility streams that are being compared). 
In this sense, the results in \citet{petri2019} have achieved significant progress. 
It is known, by now, that weak Pareto is not indispensable to get a numerical representation.

\citet{dubey2020} continue the line of investigation initiated in \citet{petri2019} but they focus on the following basic question: is it possible to strengthen the lower asymptotic Pareto condition (or, alternatively, weaken the infinite Pareto axiom to a condition stronger than lower asymptotic Pareto) while still be able to obtain a representation as in \citet{petri2019}? 
Their choice falls on the upper asymptotic Pareto axiom, which requires a stream of utilities to be ranked above another one if under the former generations belonging to a subset with positive upper asymptotic density are better off than under the latter, and no one has lower utility in the former compared to the latter.
They prove that the impossibility results in \citet{basu2003} and \citet{crespo2009} emerge again even when the infinite Pareto condition is weakened further to the upper asymptotic Pareto axiom for any non-trivial utility domain.
In their proof, the upper asymptotic density (of the subset of generations experiencing a welfare improvement) equals the maximum attainable value, namely one. 
This way, the authors demonstrate robustness of their non-existence result.

The above literature review should persuade the interested reader that the research program started in \citet{diamond1965} has been completed in the following manner. 
There is no numerical representation of social welfare orders satisfying anonymity and upper asymptotic Pareto for any non-trivial domain $Y$. 
Combining lower asymptotic Pareto with anonymity is compatible with a numerical representation for sets $Y$ with finite cardinality. 
The domain $Y$ could be enlarged, to include countably many elements having desirable order property, without forgoing the existence of an anonymous social welfare function when the Pareto condition is weakened to weak Pareto. 
Table \ref{T1} summarizes the above discussion on social welfare functions satisfying anonymity and the version of the Pareto condition which is listed in the first column.


\begin{table}[h!]
\caption{Ethical social welfare function}
\centering 
\begin{tabular}{l c l}
\hline\hline 
Efficiency axiom&$Y$& Representation  \\ [0.25ex] 
\hline
Pareto&$|Y|\geq 2$&No (\citet{basu2003})  \\
\hline
Infinite Pareto&$|Y|\geq 2$&No (\citet{crespo2009})  \\ 
\hline
Upper asymptotic Pareto&$|Y|\geq 2$& No (\citet{dubey2020})   \\ 
\hline
Lower asymptotic Pareto&$|Y|<\infty$& Yes  (\citet{petri2019}) \\ 
\hline
Lower asymptotic Pareto&$|Y|=\infty$& No (\citet{petri2019}) \\
\hline
Asymptotic density-one Pareto&$|Y|<\infty$& Yes  (\citet{petri2019}) \\
\hline
Asymptotic density-one Pareto&$|Y|=\infty$& ? \\
\hline
Weak Pareto&$Y^{\ast}$& Yes (\citet{dubey2011} \\
\hline
Weak Pareto&$Y^{\ast\ast}$& No (\citet{dubey2011} \\
\hline
Uniform Pareto&$Y$ bounded& No (\citet{sakai2016} \\
\hline\hline
\end{tabular}
\caption*{Note: $Y^{\ast}\subset\R$ not order isomorphic to the set of integers. $Y^{\ast\ast}\subset\R$ order isomorphic to the set of integers.}
\end{table}

The attentive reader will not fail to notice that the proof of the impossibility result in \citet{petri2019} can handle subsets of generations, experiencing a welfare improvement, that have a positive (but strictly less than one) lower asymptotic density.%
\footnote{His proof relies on the \emph{binary Van der Corput} sequence of rational numbers. 
For $\alpha$, $\beta\in [0, 1]$, $\alpha<\beta$, the binary Van der Corput sequence $q_n\in [0, 1]$ generates a subset of natural numbers $\{n\in\N\}$ with asymptotic density $\beta-\alpha \in (0, 1)$.} 
This is not the case for the possibility part of the characterization result in \citet{petri2019} as it continues to hold true when the subsets of generations experiencing  a welfare improvement has asymptotic density equal to one (i.e., the maximum attainable).
One could always visualize strong intuition to carry the idea of the impossibility result over to the limit, i.e., to be able to account for the subsets of generations that see a welfare improvement having lower asymptotic density equal to exactly one, even though the existing proof would be found lacking in this case.
However, progress on this front can only be achieved with a rigorous proof of the impossibility result, which is precisely what our first result takes care of.
To begin with, we name the relevant Pareto condition asymptotic density-one Pareto axiom.%
\footnote{In case the lower asymptotic density takes on the value of one, the asymptotic density exists and is equal to the lower (as well as upper) asymptotic density. 
It makes sense, therefore, to refer to this situation as asymptotic density-one.}
We show in Theorem \ref{T1} that the representation theorem in \citet{petri2019} still holds true even when the lower asymptotic Pareto condition is weakened to its limiting case, i.e., asymptotic density-one Pareto.

The issue of representability of equitable preferences remains an active field of research.
\citet{sakai2016} insists on a continuous (with respect to sup-metric) representation of ethical preferences and shows that the Pareto condition consistent with these requirements is the uniform Pareto axiom.%
\footnote{The uniform Pareto axiom requires uniform welfare gain  across all generations. 
With the addition of uniform Pareto, we have the full spectrum of the versions of the Pareto axiom as follows:
Pareto $\implies$ infinite Pareto $\implies$ upper asymptotic Pareto $\implies$ lower asymptotic Pareto $\implies$ asymptotic density-one Pareto $\implies$ weak Pareto $\implies$ uniform Pareto.
Note that each of the Pareto conditions defines a strict ranking. 
The efficiency condition based on weak preference is known as monotonicity.
Given a pair of utility sequences, if no generation is worse off in the first one than under the second, then the first  utility sequence is weakly preferred to the second.
Combining monotonicity and the Pareto condition, we can define efficient social welfare orders on $X$.}
His social welfare function also satisfies an additional axiom, namely weak non-substitution, which states that a sacrifice by the first generation that yields a uniform gain for all future generations leads to another weakly preferred utility stream.
\citet{asheim2020} impose stationarity and a condition of limited discontinuity on ethical preferences satisfying a sensitivity (Pareto) condition known as restricted dominance, and prove an impossibility-of-representation theorem.
They also conjecture that the impossibility result carries over even when limited discontinuity is dispensed with.

Our second contribution concerns the explicit construction of ethical social welfare orders.
Given the impossibility of achieving a numerical representation of equitable preferences, one way out of this problem would be either to restrict attention to consistent pair-wise ranking of utility streams, or to seek a social welfare function by getting rid of some of the axioms.
\citet[p. 1254]{svensson1980} pursued the first approach and highlighted important aspects of the continuity condition in this setting.
He observed that \enquote{in the space $X$ a continuity assumption of preferences is not only a mathematical assumption, which is often the case in finite dimensional spaces, but also reflects a value judgment}.
In view of the above, it makes sense to look at social welfare orders (or their numerical representation) satisfying the anonymity and Pareto conditions without insisting on the continuity axiom.
\citet[Theorem 2]{svensson1980} proved the existence of a social welfare order satisfying anonymity and Pareto axioms when $Y=[0,1]$. 
However, his proof relies on Szpilrajn's Lemma which is known to be a non-constructive device in the mathematical logic literature. 
Therefore, Svensson's  result requires additional scrutiny before it can be used as a guide for policy-making.

The quest for explicit description is relevant in view of the existence of ethical social welfare order established in \citet{svensson1980}.
A social welfare order allows for pairwise ranking of any pair of utility streams,  therefore it comes in handy for policy-making settings where a numerical representation of social preferences is not available.
One example, widely referenced in the literature, is the lexicographic order, which is well-known to not admit any real valued representation.
However, this order can be explicitly described so that, given any pair of utility streams, one is able to choose the preferred stream.
Therefore, we say that the lexicographic order admits explicit description or is constructive in nature.
The property of possessing a constructive nature is not a standard feature of the social welfare orders considered in the intergenerational equity literature, as the latter are shown to exist usually with the aid of the axiom of choice or equivalent tools.
Given the impossibility theorem in \citet{diamond1965} and possibility result in \citet{svensson1980}, the issue of explicit description of  ethical social welfare orders was taken up in \citet{fleurbaey2003}, keeping in mind that Svensson relied upon Szpilrajn lemma, which is a non-constructive device.
A detailed analysis in \citet{fleurbaey2003} led the authors to the following conjecture: \enquote{there exists no explicit description (that is, avoiding the axiom of choice or similar contrivances) of an ordering which satisfies the anonymity and weak Pareto axioms}.
\citet{lauwers2010} and \citet{zame2007} examined and confirmed this conjecture using well-known non-constructive devices: non-Ramsey sets and non-measurable sets, respectively.
Eventually, the investigation of the potential constructive nature of ethical social welfare orders has become a significant exercise for applications in economic policy.%
\footnote{Some of the noteworthy recent contributions on this issue are \citet{dubey2011a}, \citet{lauwers2012}, \citet{dubey2014}, \citet{dubey2014b}, \citet{lauwers2016}, \citet{dubey2016a}, and \citet{dubey2020}.}
The following table summarizes the existing results on the constructive nature of ethical social welfare orders.%
\footnote{When a social welfare function exists, an explict functional form is known, which also describes the social welfare order. 
Hence in Table \ref{T2} we report only those cases for which a numerical representation does not exist.}

\begin{table}[h!]
\caption{Construction of ethical social welfare order}
\centering 
\begin{tabular}{l c l}
\hline\hline 
Efficiency axiom&$Y$& Construction  \\ [0.25ex] 
\hline
Pareto&$|Y|\geq 2$&No (\citet{lauwers2010}, \citet{zame2007})  \\
\hline
Infinite Pareto&$|Y|\geq 2$&No (\citet{lauwers2010})  \\ 
\hline
Upper asymptotic Pareto&$|Y|\geq 2$& No (\citet{dubey2020})   \\ 
\hline
Lower asymptotic Pareto&$|Y|=\infty$& ? \\
\hline
Asymptotic density-one Pareto&$|Y|=\infty$& ? \\
\hline
Weak Pareto&$Y=[0,1]$& No (\citet{zame2007}) \\
\hline
Weak Pareto&$Y^{\ast\ast}$& No (\citet{dubey2011} \\
\hline\hline
\end{tabular}
\caption*{Note: $Y^{\ast\ast}\subset\R$ order isomorphic to the set of integers.}
\end{table}

The non-existence of ethical social welfare functions satisfying asymptotic density-one Pareto raises the  question of whether the social welfare order under consideration (which exists in view of \citet[Theorem 2]{svensson1980} in the case of ethical social welfare orders satisfying the Pareto axiom) is constructive or, in other words, admits explicit description.
A negative answer would also apply to the impossibility result in \citet{petri2019} which deals with the lower asymptotic Pareto case.%
\footnote{In \citet[Table 1, p.18]{dubey2020} we have flagged this up as an open question.}
We examine this question and, in Theorem \ref{T2}, we show that an ethical social welfare order satisfying the asymptotic density-one Pareto axiom is constructive in nature if and only if the cardinality of the utility domain is finite.
Therefore, we provide a characterization of the constructive nature which can be useful for policy purposes.

The remainder of the paper is organized as follows. 
Section \ref{sec:2} gathers some preliminary concepts, notations, a brief description of the notion of asymptotic density, the definitions of relevant binary relations  and the equity and efficiency axioms imposed on the social welfare orders under examination. 
In section \ref{sec:3}, we state and prove our results on the representation of social welfare orders satisfying the anonymity and asymptotic density-one Pareto axioms.
Section \ref{sec:4} contains our result on the constructive nature of social welfare orders satisfying the anonymity and asymptotic density-one Pareto axioms.
Section \ref{sec:5} concludes with an open question which is left for future research.
The proofs are relegated to the Appendix. 

\section{Preliminaries}

\label{sec:2}

Let $\R$, $\Q$, and $\N$ be the set of real numbers, rational numbers, and natural numbers, respectively. 
For all $y$, $z\,\in \,\R^{\N}$, we write $y\geq z$ if $y_{n}\geq z_{n}$, for all $n\in \N$; $y>z$ if $y\geq z$ and $y\neq z$; and $y\gg z$ if $y_{n}>z_{n}$ for all $n\in \N$. 
Given any $x\in \R^{\N}$and $N\in \N$, we denote the vector consisting of the first $N$ components of $x$ by $x(N)$, and the tail sequence of $x$, from the element $N+1$ onward, by $x[N]$. 
Formally, $x(N)=\left(x_{1}, x_{2}, \cdots, x_{N}\right)$ and $x[N]=\left(x_{N+1}, x_{N+2},\cdots \right)$.

It is useful to recall the definition of asymptotic density of any $S\subset \N$. 
As usual, let $|\cdot|$ denote the cardinality of a given finite set. 
The lower asymptotic density of $S$ is defined as follows: 
\begin{equation*}
\underline{d}(S)=\underset{n\rightarrow \infty }{\liminf }\;\frac{|S\cap \{1, 2, \cdots, n\}|}{n}.
\end{equation*}%
Similarly, the upper asymptotic density of $S$ is defined as follows: 
\begin{equation*}
\overline{d}(S)=\underset{n\rightarrow \infty }{\limsup }\;\frac{|S\cap \{1, 2, \cdots, n\}|}{n}.
\end{equation*}
Note that for any set $S$ with $|S|<\infty$, $\overline{d}(S)=\underline{d}(S)=0$.
However, in general $\overline{d}(S)$ and $\underline{d}(S)$ need not coincide as is the case for $S = \bigcup_{k=1}^{\infty} \{(2k-1)!, (2k)!\}$, where $\overline{d}(S)=1$ and $\underline{d}(S)=0$.
One says that $S$ has asymptotic density $d(S)$ if $\underline{d}(S)=\overline{d}(S)$, in which case $d(S)$ is equal to this common value.
Formally, 
\begin{equation*}
d(S)=\underset{n\rightarrow \infty }{\lim }\;\frac{|S\cap \{1, 2, \cdots, n\}|}{n}.
\end{equation*}
For $S:= \{n!:n\in\N\}$, we have that $d(S)=0$ and  $d(\N\setminus S)=1$.
Given $A$, $B\subset\N$, let $A\;\Delta\; B:= (A\setminus B)\cup (B\setminus A)$. 
If $|A\;\Delta\; B|<\infty$, then it turns out that $\underline{d}(A)=\underline{d}(B)$, $\overline{d}(A)=\overline{d}(B)$ and $d(A)=d(B)$.

\subsection{Social welfare relations}

Let $Y\subset \R$ be the set of all possible utilities that any generation can achieve. 
Then, $X\equiv Y^{\N}$ is the set of all feasible utility streams. 
We denote an element of $X$ by $x$, or alternately by $\langle x_{n}\rangle $, depending on the context. 
If $\langle x_{n}\rangle \in X$, then $\langle x_{n}\rangle =\left( x_{1},x_{2},\cdots\right)$. 
For all $n\in \N$, $x_{n}\in Y$ represents the amount of utility that period-$n$ generation earns.

A \emph{social welfare relation} is a binary relation $\succsim$ on $X$ which is reflexive and transitive. 
Its symmetric and asymmetric parts, denoted by $\sim$ and $\succ$, respectively, are defined in the usual manner. 
Thus, we write $x \sim y$ when $x\succsim y$ and $y\succsim x$ both hold, and we write $x\succ y$ when $x\succsim y$ holds, but $y\succsim x$ does not hold.
A \emph{social welfare order} is, by definition, a complete and transitive binary relation on $X$. 
Given a social welfare order $\succsim$, one says that $\succsim $ can be \emph{represented} by a real-valued function, called a \emph{social welfare function}, if there is a mapping $W:X\rightarrow \R$ such that for all $x$, $y\in X$, $x\succsim y$ if and only if $W(x)\geq W(y)$.

\subsection{Equity and efficiency axioms}\label{sec:3}

We will be dealing with the following equity and efficiency axioms we may want the social welfare relations to satisfy.

\begin{definition}
\label{D1} \emph{Anonymity: If $x, y\in X$, and there exist $i, j\in \N$ such that $y_{j} = x_{i}$ and $x_{j} = y_{i}$, while $y_{k} = x_{k}$ for all $k\in \N\setminus \{i, j\}$, then $x\sim y$.}
\end{definition}

\begin{definition}
\label{D2} \emph{Lower asymptotic Pareto: Given $x, y\in X$, if $x\geq y$ and $x_{i}>y_{i}$ for all $i$}$\in $\emph{$S\subset \N$ with $\underline{d}(S)>0$, then $x\succ y$.}
\end{definition}

\begin{definition}
\label{D3} \emph{Asymptotic density-one Pareto: Given $x,y\in X$, if $x\geq y$ and $x_{i}>y_{i}$ for all $i$}$\in $\emph{$S\subset \N$ with $d(S)=1$, then $x\succ y$.}
\end{definition}
Anonymity (or the Pareto condition) alone permits consistent evaluation by a real-valued function, e. g., $W(x) = \underset{t\rightarrow \infty}{\liminf}\; x_t$ $\left(W(x) = \sum_{t=1}^{\infty} \delta^{t-1}x_t,\;\text{with}\;\delta\in(0, 1)\right)$.
A conflict occurs when we look to combine anonymity and Pareto (or any of its numerous weaker versions) and demand a real-valued representation of the resulting ethical social welfare relation.
The Pareto condition considered in this paper is asymptotic density-one Pareto.
Lower asymptotic Pareto implies asymptotic density-one Pareto but the converse is not true. 
Hence, asymptotic density-one Pareto is strictly weaker than lower asymptotic Pareto.

In order to explain the thrust of our contribution, we briefly discuss the conflict between the notions of equity and efficiency and their consistent evaluations through social welfare relations, social welfare orders and social welfare functions.
An example of an ethical social welfare relation is as follows:
\begin{example}\label{Ex01}
\emph{\textbf{Suppes-Sen grading principle}: $x\succcurlyeq_{S} y$ if and only if there exists a permutation $\pi$ of $\N$,
\footnote{$\pi:\N\rightarrow \N$ is a bijection such that there exists $N\in\N$, $\pi(n)=n$ for all $n>N$.} such that $x(\pi)\geq y$.
It satisfies both the anonymity and Pareto conditions.%
\footnote{It is not a complete binary relation as there are pair of sequences that cannot be ranked. 
Take, for example, $x=\{1, 0, 1, 0, \cdots\}$ and $y=\{0, 1, 0, 1, \cdots\}$.}.}
\end{example}
It is important to point out that not all equity axioms are amenable to consistent evaluation when combined with efficiency criteria. 
\citet[Theorem 5.1]{petri2019} characterizes the equity (as a variation of anonymity) and efficiency conditions for which no social welfare relation exists.%
\footnote{Non-existence of a binary relation satisfying a different class of equity condition, i.e., the generalized Pigou-Dalton transfer principle, is established in \citet{dubey2020a}.} 
\citet{svensson1980} extended the Suppes-Sen grading principle, using Szpilrajn lemma, to obtain a complete ethical binary relation, i.e., an ethical social welfare order.
To explain its non-constructive nature, consider the lexicographic order on infinite utility sequences, defined below:
\begin{example}\label{Ex02}
\emph{\textbf{Lexicographic order}: $x\succ_L y$ if and only if there exists $T\in\N$ such that  $x_t=y_t$ for all $t<T$ and $x_T>y_T$.
It is a complete binary relation, hence it is an order which admits explicit description as described above.}
\end{example}
The conjecture in \citet{fleurbaey2003}, confirmed by \citet{lauwers2010} and \citet{zame2007}, demonstrates that Paretian ethical social welfare order (which is shown to exist in \citet{svensson1980}) does not admit explicit description, therefore it is non-constructive in nature.

\section{Social welfare functions satisfying asymptotic density-one Pareto and anonymity}\label{sec:3} 
In this section we examine the representability of social welfare orders satisfying the asymptotic density-one Pareto and anonymity axioms.

\subsection{A possibility result}
Lemma \ref{L0} presents a procedure for explicitly constructing a social welfare function satisfying asymptotic density-one Pareto and anonymity, if the cardinality of $Y$ is finite, i. e., $|Y|<\infty$.

\begin{lemma}\label{L0}[\citet[p. 860]{petri2019}] 
Let $X=Y^{\N}$, $|Y|<\infty$.
There exists a social welfare function satisfying the anonymity and asymptotic density-one Pareto axioms.
\end{lemma}

Let $X=Y^{\N}$, $|Y|<\infty$, and let $W:X\rightarrow \R$ be defined as follows:

\begin{equation}  \label{ExEq1}
W(x):=\underset{n \rightarrow \infty}{\liminf}\;\;\frac{\sum_{k=1}^n x_k}{n}.
\end{equation}

\noindent Next, define the following binary relation $\succsim $ \emph{on} $X$:

\begin{equation}
x\succsim y\quad \text{if and only if}\quad W(x)\geq W(y).  \label{ExEq2}
\end{equation}

\noindent Clearly, (\ref{ExEq1}) is a social welfare function that represents the social welfare order given by (\ref{ExEq2}). 
\citet[p. 860]{petri2019} has shown that the latter satisfies anonymity and lower asymptotic Pareto.
Since asymptotic density-one Pareto is weaker than lower asymptotic Pareto, the social welfare order defined by (\ref{ExEq2}) also satisfies asymptotic density-one Pareto  and anonymity, when $|Y|<\infty$. 

\begin{remark}\label{R1}
\emph{Since (\ref{ExEq1}) is an explicit formula for the social welfare function, the underlying social welfare order (\ref{ExEq2}) can be explicitly described, i.e., it is a constructive object.}
\end{remark}

\subsection{An impossibility result}\label{s4.1}
Given Lemma \ref{L0}, we only need to consider a set $Y$ with  $|Y|=\infty$ to deal with non-existence of a representation. 
Recall that a set $S$ with $|S|=\infty$, contains either a subset which is order isomorphic to $\N (<)$, or a subset order isomorphic to the set of negative integers.%
\footnote{Pick $s_1\in S$, since $S$ is non-empty, and set $s_{k+1}>s_k$ for all $k\in\N$ recursively (which is  possible since $|S|=\infty$), to obtain the subset $\{s_k:k\in\N\}$ which is order isomorphic to $\N(<)$.
In case $s_K$ turns out to be the maximal element of $S$ for some $K\in\N$, then we consider the sequence of decreasing elements, $s_{k+1}<s_k$ for all $k\in\N$ (which must exist since $|S|=\infty$) to obtain the desired subset order isomorphic to the set of negative integers.}
It is known from \citet[Proposition 2]{dubey2011} that the existence result of ethical social welfare functions satisfying Pareto axiom is invariant to monotone  transformations of the domain $Y$.
Therefore, it suffices to give a proof for the case $Y=\N$, or $Y$ equal to the set of negative integers. 
Lemma \ref{L1} shows that no social welfare function  satisfying asymptotic density-one Pareto and anonymity exists if $Y=\N$.
The other case of $Y$ equal to the set of negative integers can be dealt with using similar arguments and is left to the reader.

\begin{lemma}\label{L1} 
There does not exist any social welfare function satisfying asymptotic density-one Pareto and anonymity on $X=Y^{\N}$, with $Y=\N$.
\end{lemma}
Lemma \ref{L1} could be applied to establish the incompatibility of numerical representation, asymptotic density-one Pareto and anonymity.
\begin{example}
\emph{\begin{enumerate}[(a)]
\item{Let $Y:= \left\{\frac{1}{n}: n\in \N\right\}\subset[0, 1]$. 
Since the set $Y$ is order isomorphic to the set of negative integers, no social welfare function exists.}
\item{Let $Y:= \left\{\frac{n}{1+n}: n\in \N\right\}\subset[0, 1]$. 
Since the set $Y$ is order isomorphic to $\N(<)$, no social welfare function exists.}
\item{Let $Y:= \Q\cap [0, 1]$. 
Since $Y$ contains the subset $\left\{\frac{n}{1+n}: n\in \N\right\}\subset[0, 1]$ which is order isomorphic to $\N(<)$, no social welfare function exists.}
\end{enumerate}}
\end{example}

\noindent Combining Lemma \ref{L0} (and the related remarks) with Lemma \ref{L1}, yields the following theorem.
The proof is straightforward and is omitted.

\begin{theorem}
\label{T1} There exists a social welfare function satisfying asymptotic density-one Pareto and anonymity on $X=Y^{\N}$ if and only if $|Y|<\infty$.
\end{theorem}

\section{Construction of social welfare orders satisfying asymptotic density-one Pareto and anonymity}\label{sec:4}

Let $A$ be an infinite set and, for $n\in \N$, let $[A]^n$ be the collection of all subsets of $A$ with exactly $n$ elements. 
\citet[Theorem A]{ramsey1928b} shows that for each subset $\mathscr{A}$ of $[A]^n$, there exists an infinite set $B\subset A$ such that either $[B]^n\subset \mathscr{A}$ or $[B]^n\cap \mathscr{A}=\emptyset$.
$[B]^n$ is called a Ramsey collection of sets.

Ramsey's theorem fails to hold when $n$ is replaced by $\N$, i.e., the set of natural numbers. 
In other words, there exists a subset $\mathscr{C}$ of $[A]^{\N}$ such that for each infinite subset $S$ of $A$ the class $[S]^{\N}$ intersects both $\mathscr{C}$ and its complement $[A]^{\N}\setminus \mathscr{C}$.
Such a set $\mathscr{C}$ is said to be a non-Ramsey collection of sets.
The formal definition of non-Ramsey collection of sets is as follows:

Let $T$ be an infinite subset of $\N$. 
We denote by $\Omega (T)$ the collection of all infinite subsets of $T$, and we let $\Omega$ denote the collection of all infinite subsets of $\N$.

\begin{definition}
\label{D5} \emph{A collection of sets $\Gamma \subset \Omega $ is said to be \emph{non-Ramsey} if for every $T\in \Omega $, the collection $\Omega (T)$ intersects both $\Gamma $ and its complement $\Omega \diagdown \Gamma $.}
\end{definition}

A non-Ramsey collection of sets is a non-constructive object, which means that its existence cannot be proven without using the axiom of choice.%
\footnote{This result has been proven in \citet{mathias1970}. 
Such a result, together with the analogous results on Lebesgue measurable and Baire sets proven in \citet{solovay1970}, is considered a corner-stone in forcing theory and descriptive set theory.}
If the existence of an ethical social welfare order implies the existence of a non-Ramsey collection of sets, the social welfare order at hand is considered to be non-constructive in nature.%
\footnote{This is because the existence of a non-Ramsey collection of sets (a non-constructive object) is necessary for the existence of the ethical social welfare order at hand.}

In light of Remark \ref{R1}, we need to examine the constructive nature of ethical social welfare orders satisfying asymptotic density-one Pareto when $|Y|=\infty$.
Also, it will suffice to prove the result for $Y=\N$ or $Y$ being the set of negative integers (as per the discussions preceding the statement of Lemma \ref{L1}). 
In Lemma \ref{L2} below we assume that $Y=\N$ and show that the existence of a social welfare order satisfying anonymity and asymptotic density-one Pareto implies the existence of a non-Ramsey collection of sets.


\begin{lemma}
\label{L2} Let $Y=\N$, and assume that there is a social welfare order on $X=Y^{\N}$ satisfying the anonymity and asymptotic density-one Pareto axioms. 
Then, there exists a non-Ramsey collection of sets.
\end{lemma}

\noindent Combining Lemma \ref{L0} with Lemma \ref{L2} leads to the following theorem.
Its proof is straightforward and is omitted.

\begin{theorem}
\label{T2} There exists a constructive social welfare order satisfying the anonymity and asymptotic density-one Pareto axioms on $X=Y^{\N}$ if and only if $|Y|<\infty$.
\end{theorem}

\section{Concluding Remarks}\label{sec:5} 
In this paper we have focused on social welfare orders, on infinite utility streams, satisfying the anonymity and asymptotic density-one Pareto axioms. 
We have characterized the restriction on the domain $Y\subset\R$ under which an ethical social welfare order is representable and admits explicit construction.
The necessary and sufficient conditions (imposed on the set $Y$) for the existence of a social welfare function are identical to the condition (on $Y$) for the social welfare order under examination to be constructive.
Thus, either there exists a social welfare function (with an explicit formula), or the social welfare order at hand is non-constructive.
From the perspective of a policy maker, the existence of ethical social welfare orders does not provide any actionable information if no social welfare function exists.

The issue of aggregating infinite utility streams into a social welfare relation that respects certain equity and efficiency principles is an active field of research. 
An open question we plan to investigate in the future is the representability and construction of social welfare relations satisfying the anonymity and an \enquote{almost weak Pareto} condition.
Specifically, according to the almost weak Pareto condition, in order for the utility stream $x$ to be ranked socially preferred to $y$, all but finitely many generations must be better off in $x$ than in $y$.%
\footnote{Note that this axiom is weaker than asymptotic density-one Pareto and stronger than weak Pareto.}

\section{Appendix}

\begin{proof}[\textbf{Proof of Lemma} \ref{L1}]

Suppose, by way of obtaining a contradiction, that $W: X\rightarrow \R$ is a social welfare function satisfying anonymity and asymptotic density-one Pareto.
The gist of the argument used in the proof can be summarized as follows.
Using asymptotic density-one Pareto, for each $r\in(0, 1)$ we construct a pair of sequences $x(r)$ and $z(r)$ such that $W(z(r))>W(x(r))$.
Also for $s>r$, using anonymity and asymptotic density-one Pareto we get that $x(s)$ is such that $W(x(s))>W(z(r))$.
This construction implies the existence of a non-empty interval $[W(x(r)), W(z(r))]$, for each $r\in(0, 1)$, disjoint from the interval $[W(x(s)), W(z(s))]$ for $s\in(0, 1)$, $s\neq r$.
This enables us to pick a rational number $q(r)$ in the interval $[W(x(r)), W(z(r))]$ for each $r\in(0, 1)$ which leads to a contradiction since $\Q$ is countable whereas $(0, 1)$ is uncountable.

Now, let $q_1$, $q_2$, $\cdots$ be an enumeration of rational numbers in $(0, 1)$.
Let $r\in (0, 1)$, and for each $k\in\N$ let $u_1(r): = \min \left\{n \in \N: q_n \geq r\right\}$ and  $u_k(r): = \min \left\{n \in \N\setminus \left\{u_l(r): l<k\right\}: q_n \geq r\right\}$.
Let $U(r):= \{(u_k(r))!: k\in\N\}$ and $L(r):= \N\setminus U(r) = \{l_k(r): l_k(r)<l_{k+1}(r), k\in\N\}$.
Note that $d(U(r))=0$ and $d(L(r))=1$.
Define the utility stream  $\langle x(r)\rangle$ as follows:%
\begin{equation}\label{03}
x_t(r) = \left\{ 
\begin{array}{ll}
1& \text{if}\; t\in U(r),\\
m+1& \text{if}\; t= l_m(r), m\in \N.%
\end{array}
\right.
\end{equation}
Utility stream $\langle z(r)\rangle$ is defined in an identical fashion by taking $U(r)\setminus \{(u_1(r))!\}$, and its complement, $L(r)\cup \{(u_1(r))!\}$:
\begin{equation}\label{04}
z_t(r) = \left\{ 
\begin{array}{ll}
1& \text{if}\; t\in U(r)\setminus \{(u_1(r))!\},\\
m+1& \text{if}\; t= l_m(r) \in L(r)\cup \{(u_1(r))!\}, m\in \N.%
\end{array}
\right.
\end{equation}
Then, $z(r)>x(r)$ since $z_{(u_1(r))!}(r)>1=x_{(u_1(r))!}(r)$; $\forall k < (u_1(r))!$ $z_k(r)=x_k(r)$; $\forall k \in U(r)$, $k > (u_1(r))!$ $z_k(r)=1=x_k(r)$; and $\forall k \in L(r)$, $k > (u_1(r))!$ $z_k(r) > x_k(r))$.
Let $S:= \overset{\infty}{\underset{k=1}{\bigcup}} \N\cap \left((u_k(r))!, (u_{k+1}(r))!\right)$.
Then, for all $t\in S$, $z_t(r)>x_t(r)$ and $d(S)=1$.
Hence, $z(r)\succ x(r)$ by asymptotic density-one Pareto, therefore
\begin{equation}\label{P1Ea}
W(z(r))>W(x(r)).
\end{equation}
For $s\in (r, 1)$,  $U(s)\subset U(r)$. 
Let  $U(r, s):= U(r)\setminus U(s)$. 
Then $|U(s)|= |\{n!\in\N: q_n\geq s\}|=\infty$ and $|U(r, s)| =|\{n!\in\N: r<q_n<s\}|=\infty$.
Consider $\langle x(s)\rangle$ and $\langle z(s) \rangle$ as defined by (\ref{03}) and (\ref{04}), respectively.
Let $u_1:= \min \left\{n!: n!\in U(r, s)\right\}$ and $u_2:= \min \left\{n!: n!\in U(r, s)\setminus \{u_1\right\}\}$.
There are two cases to consider.
\begin{enumerate}[]
\item{\underline{Case (a)}: }{$u_1= (u_1(r))!$, i.e., $q_{u_1(r)}\in [r, s)$. 
Then $x(s)>z(r)$, since $x_{u_2}(s)> 1=z_{u_2}(r)$, $\forall k < u_2 (x_k(s)=z_k(r))$,  $\forall k \in U(s) (x_k(s)=1=z_k(r))$, and $\forall k \in L(s)(k > u_2 \Rightarrow x_k(s) > z_k(r))$.
Since $d\left(\{n\in L(s): n > u_2\}\right)=1$, $x(s) \succ z(r)$ by asymptotic density-one Pareto.}
\item{\underline{Case (b)}: }{$u_1 > (u_1(r))!$, i.e., $q_{u_1(r)}\in [s, 1)$.
Since $z_t(r)>x_t(s)$ for all coordinates $t\in [(u_1(r))!, u_1)\setminus U(r)$, $x(s)\not> z(r)$.%
\footnote{However, performing a finite permutation $\pi$ on $z(r)$, using asymptotic density-one Pareto, one obtains $z^{\pi}$ such that $x(s)>z^{\pi}$ and $x(s)\succ z^{\pi}$.
Since by anonymity $z^{\pi}\sim z(r)$, we get $x(s)\succ z(r)$.}
Comparing $x(s)$ and $z(r)$, we realize that the following additional inequalities hold: $\forall t<(u_1(r))!$, $z_t(r) = x_t(s)$, $\forall t\in \left((u_1(r))!, u_1\right)\cap U(r)$, $z_t(r) = x_t(s) = 1$, and $z_{u_1}(r)=1< x_{u_1}(s)$, $\forall t\in \left(u_1, u_2\right)$, $z_t(r) = x_t(s)$, and $z_{u_2}(r)=1<  x_{u_2}(s)$, $\forall t>u_2$, $t\in L(s)$,  $x_t(s)>z_t(r)$, and $\forall t>u_2$, $t\in U(s)$,  $x_t(s)=z_t(r)=1$.
Let $\{a_1, a_2, \cdots a_K\}$ be an increasing enumeration of $[(u_1(r))!, u_1] \cap L(r)$ and apply the following finite permutation $\pi$ to $z(r)$ to obtain $z^{\pi}$:
\begin{equation}
\pi(t):=
\begin{cases}
t \quad &\text{if} \quad t \notin [(u_1(r))!, u_1]\\ 
u_{1}\quad &\text{if}\quad  t = (u_1(r))!\\
a_{j-1} \quad &\text{if}\quad t=a_j \text{ (for $j \geq 2$)}.
\end{cases}
\end{equation}
Then, $z^{\pi}_{(u_1(r))!} = z_{u_1}(r) =1$, $\forall t\in ((u_1(r))!, u_1] \cap U(r)$, $^{\pi}_{t} = z_{t}(r)$, $z^{\pi}_{t} = z_{t^{\prime}}(r)$, where $t^{\prime}$ and $t$ occupy the same position in the increasing enumerations of the sets  $[(u_1(r))!, u_1) \cap L(r)$ and $((u_1(r))!, u_1] \cap L(r)$ respectively, $z^{\pi}_t =  z_{t}(r)$ for the remaining $t\in \N$.
Hence, $z^{\pi}\sim z(r)$, and, by anonymity, $W(z^{\pi}) = W(z(r))$.
Also  $x(s)>z^{\pi}$, $\forall t<u_2$, $x_t(s)\geq z^{\pi}_t$, $x_{u_2}(s)>1=z^{\pi}_{u_2}$, $\forall t>u_2$, $t\in L(s)$,  $x_t(s)> z^{\pi}_t$, and $\forall t>u_2$, $t\in U(s)$,  $x_t(s)=z^{\pi}_t=1$.
Since $d\left(\{n \in L(s): n > u_2\}\right)=1$, $x(s)\succ z^{\pi}$ by asymptotic density-one Pareto, and $W(x(s)) > W(z^{\pi})$.
Combining $z(r)\sim z^{\pi}$ and $z^{\pi}\prec x(s)$, we get $z(r) \prec x(s)$.%
\footnote{To fix ideas, let $\{1!, 2!, 3!, 4!, 7!\}\in U(r)$ and $\{1!,  2!, 7!\} \in U(s)$.
Then $u_1(r)!=1!$, $u_1=3!$, $u_2=4!$, $x(r)= \{{\color{blue}{1}}, {\color{blue}{1}}, 2, 3, 4, {\color{blue}{1}}, 5, \cdots, 21, {\color{blue}{1}}, 22, \cdots\}$, $z(r)= \{2, {\color{blue}{1}}, 3,  4, 5, {\color{blue}{1}}, 6, \cdots, 22, {\color{blue}{1}},  23, \cdots\}$ and $x(s)= \{{\color{blue}{1}}, {\color{blue}{1}}, 2, 3, 4, {\color{red}{5}}, 6, \cdots, 22, {\color{red}{23}}, 24, \cdots\}$.
Though $z_t(r)>x_t(s)$, for $t\in \{1, u_1-1\}\setminus \{u_1(r)!, u_2(r)!\}$, by a finite permutation of coordinates of $z(r)$, we get $z^{\pi} =  \{{\color{blue}{1}}, {\color{blue}{1}}, 2, 3,  4, {\color{red}{5}}, 6, \cdots, 22, {\color{blue}{1}},  23, \cdots\}$. 
For all $t< u_2$, $x_t(s)=z^{\pi}_t$, $x_{u_2}(s)=23>1=z^{\pi}_{u_2}$ and for all $t\in L(s)$, $t\geq u_2$, $x_t(s)>z^{\pi}_t$.}}
\end{enumerate}
In both cases,
\begin{equation}\label{P1Ec}
W(z(r)) < W(x(s)).
\end{equation}
Therefore,  (\ref{P1Ea}) and (\ref{P1Ec}) imply that $(W(x(r)), W(z(r))$ and $(W(x(s)), W(z(s))$ are non-empty and disjoint open intervals.    
Hence, because $r$ and $s$, with $r<s$, were arbitrary, by density of $\Q$ in $\R$ we obtain a one-to-one mapping from $(0, 1)$ to $\Q$ which is impossible as the latter set is countable.
\end{proof}

In Lemma \ref{L2}, an arbitrary infinite sequence $T$ (with infinite complement) in $\N$ will be used to define a pair of utility streams, namely $x(T)$ and $Y(T)$.
Take the collection of sets $T\in \N$ such that $x(T)\prec y(T)$ as a candidate for non-Ramsey collection of sets.
The completeness property of the ethical social welfare relation yields three possible rankings, namely $x(T)\prec y(T)$, $x(T)\sim y(T)$ and $x(T)\succ y(T)$ for the infinite sequence $T$.
Thus, either $T$ belongs to this collection, i.e., $x(T)\prec y(T)$ or it does not, i.e., $x(T)\sim y(T)$ or $x(T)\succ y(T)$.
In case $T$ belongs to the collection,  there exists a subsequence $S$ of $T$ such that $S$ does not belong to the collection, i.e., $x(S)\succ y(S)$.
If $T$ does not belong to the collection, we can find a subsequence $S$ of $T$ such that $S$ belongs to the collection, i.e., $x(S)\prec y(S)$.
Anonymity and the asymptotic density-one Pareto ranking will play a role in the proof of this fact.
Moreover, we will establish that the candidate collection of sets is in fact a non-Ramsey set.

The asymptotic density of the following subset of $\N$ will be useful for the proof of Lemma \ref{L2}. 
Let $N:=\left\{n_k: k\in\N, n_k<n_{k+1}\right\}$ be an arbitray sequence, $U_k(N) := \N\cap \left((n_{2k-1})!, (n_{2k+1})!-\frac{(n_{2k+1})!}{(n_{2k})!}\right]\neq \emptyset$ for $k\in\N$ and 
$U(N) := \overset{\infty}{\underset{k=1}{\bigcup}}U_k(N)$.
Since $\left|U_k(N)\right|= (n_{2k+1})!-\frac{(n_{2k+1})!}{(n_{2k})!} - (n_{2k-1})!$,
\begin{align*}
\underline{d}\left(U(N)\right) &= \underset{m\rightarrow \infty}{\liminf} \frac{\left|\overset{m}{\underset{k=1}{\bigcup}} U_k(N)\right|}{(n_{2m+1})!} \geq  \underset{m\rightarrow \infty}{\lim} \frac{\left| U_m(N)\right|}{(n_{2m+1})!}= \underset{m\rightarrow \infty}{\lim} 1- \frac{1} {(n_{2m})!}-\frac{(n_{2m-1})!} {(n_{2m+1})!}=1,
\end{align*}
$d\left(U(N)\right)=1$, and $d\left(\N\setminus U(N)\right)=0$.
The following inequality regarding the cardinality of subsets of $\N$ will be used in the proof of Lemma \ref{L2}.

\begin{equation}\label{L2E1}
\frac{t_{2m+2}!}{t_3!}>\frac{t_{2m+2}!}{t_{2m+1}!} + \cdots+ \frac{t_4!}{t_3!}\;\text{for all}\; m>1.
\end{equation}
Since there are $m$ terms on the right hand side in (\ref{L2E1}), the above inequality is equivalent to:
\begin{align}\label{L2E2}
\left(\frac{t_{2m+2}!}{t_3!}-m\frac{t_{2m+2}!}{t_{2m+1}!}\right)+\left(\frac{t_{2m+2}!}{t_3!}-m\frac{t_{2m}!}{t_{2m-1}!}\right) + \cdots+ \left(\frac{t_{2m+2}!}{t_3!}-m \frac{t_4!}{t_3!}\right)&>0\notag\\
\frac{t_{2m+2}!}{t_{2m+1}!}\left(\frac{t_{2m+1}!}{t_3!}-m\right)+\frac{t_{2m}!}{t_{2m-1}!}\left(\frac{t_{2m+2}!}{t_{2m}!}\frac{t_{2m-1}!}{t_3!}-m\right) + \cdots+ \frac{t_4!}{t_3!}\left(\frac{t_{2m+2}!}{t_4!}-m \right)&>0.
\end{align}
Since $\frac{t_{2m+2}!}{t_{2m+1}!}>1$, $\frac{t_{2m}!}{t_{2m-1}!}>1$, $\cdots$, $\frac{t_4!}{t_3!}>1$, it suffices to show that the expressions inside each of the $m$ parentheses  in (\ref{L2E2}) are positive.
For the terms inside the first parenthesis,  $t_{2m+1}\geq t_3+(2m-2)$, $t_3\geq 3$, and
$\frac{t_{2m+1}!}{t_3!}\geq \frac{(t_3+2m-2)\cdots(t_3+1) t_3!}{t_3!}$ $\geq (3+2m-2)\cdots(3+1)= (2m+1)\cdots(4)>m$.
As to the expressions inside the second parenthesis, $\frac{t_{2m-1}!}{t_3!}\geq 1$ and $\frac{t_{2m+2}!}{t_{2m}!} \geq \frac{(t_{2m}+2)!}{t_{2m}!}$ $= \frac{(t_{2m}+2) (t_{2m}+1) t_{2m}!}{t_{2m}!} \geq (2m+2)(2m+1) > m$.
To complete the proof, one can just use similar arguments for the expressions inside the remaining parentheses.

\begin{proof}[\textbf{Proof of Lemma} \ref{L2}]
Given the sequence $N$ defined above, let $U(N):=\left\{u_k: u_k<u_{k+1}, \; \text{for all}\; k\in \N\right\}$.
Let utility stream $\langle x(N)\rangle$ be defined as follows:  
\begin{equation}
x_{t} (N)=
\begin{cases}
1 \quad &\text{if}\quad  t\in  \N\setminus U(N),\\ 
k+1 \quad &\text{if}\quad  t = u_k, u_k\in U(N)\; \text{for}\; k\in \N,
\end{cases}
\end{equation}
and define the utility stream $\langle y(N)\rangle:= \langle x(N\setminus \{n_1\})\rangle$.
Let $\succsim $ be a social welfare order satisfying anonymity and asymptotic density-one Pareto. 
We show that the collection of sets $\Gamma \equiv \{N\in \Omega: x(N)\prec y(N)\}$ is non-Ramsey, i. e., $\forall T\in \Omega$, the collection $\Omega (T)$ intersects both $\Gamma $ and $\Omega\diagdown \Gamma$.
To this end, it is enough to show that for every $T\in \Omega$, (1) if $T\in \Gamma$ then there exists $S\in \Omega(T)$ such that $S\notin \Gamma$, and, (2) if $T\notin \Gamma$, then there exists $S\in \Omega(T)$ such that $S\in \Gamma$.
As the binary relation is assumed to be complete, there are three cases to consider. 
\begin{enumerate}[(a)]
\item{$x(T)\prec y(T)$ or $T\in \Gamma$:
Take $S := T\setminus \{t_1\} = \{t_2, t_3, t_4,\cdots\} \in \Omega(T)$.
Then $y_t(T) = x_t\left(T-\{t_1\}\right)= x_t\left(S\right)\;\forall\; t\in \N$, and $y(T) \sim x(S)$.
Since $y_t(S)= x_t\left(T-\{t_1, t_2\}\right)$, $x_t(T)\geq y_t(S)$ for all $t\in \N$, and  $x_t(T)>y_t(S)\geq 1$, for all $t \in U(T)$.%
\footnote{To fix ideas, let $T=\N$.
Then $x(T) = \{1, 2, 3, 1, 1, 1, 4, \cdots, 112, 1, 1, 1, 1, 1, \cdots\}$, $y(T) = \{1, 1, 2, 3, \cdots, 19, 1, 1, 1, 1, 20,  \cdots\} = x(S)$, $y(S) = \{1, 1, 1, 1, 1, 1, 2, 3, \cdots, 110, 1, 1, 1, 1, 1, \cdots\}$ hence $x_t(T)>y_t(S)$ for all $t\in U(T)$.}
Given $d\left(U(T)\right) = 1$, applying asymptotic density-one Pareto, we get $y(S) \prec x(T)$.
Thus $y(S) \prec x(T) \prec y(T) \sim x(S)$. By trnasitivity $y(S) \prec x(S)$. 
Hence, $S\notin \Gamma$.}
\item{$y(T)\prec x(T)$ or $T\notin \Gamma$: 
Drop $t_1$ and $t_{4}$, $t_{5}$,  $\cdots$, $t_{2m}$, $t_{2m+1}$ (with $m \geq 2$) so as to obtain $S :=\{t_2, t_3, t_{2m+2}, t_{2m+3}, \cdots\}\in \Omega(T)$ such that $\left|U_1(T)\right|=(t_3)!-\frac{(t_3)!}{(t_2)!} -(t_1)! < \frac{(t_5)!}{(t_4)!}+ \cdots+ \frac{(t_{2m+1})!}{(t_{2m})!}$.

For $y(T)$ and $x(S)$, $y_t(T)=x_t(S)\geq 1$ for all $t\in\left[1, t_4!-\frac{t_4!}{t_3!}\right]$; in view of (\ref{L2E1}), for the coordinates $t\in \left[t_{2}!, t_{2m+2}!\right]\cap\N$, $\left|\left\{t: y_t(T)=1<x_t(S)\right\}\right| < \left|\left\{t: y_t(T)>x_t(S)\geq 1\right\}\right|$, $y_t(T) - x_t(S)\geq 1$ for all $t\in\overset{\infty}{\underset{k=m+1}{\bigcup}}\left(t_{2k}!, t_{2k+2}!- \frac{t_{2k+2}!}{t_{2k+1}!}\right]\cap\N$, and $y_t(T)=x_t(S)=1$ for the remaining $t\in \N$.
Thus, $x_t(S)\leq y_t(T)$ for all but finitely many $t\in \N$.
Next, permute the coordinates of $x(S)$ in $\left[t_{2}!, t_{2m+2}!\right]$  such that  $y_t(T)=1<x_t(S)$ with equally many elements from the remaining coordinates in $\left[t_{2}!, t_{2m+2}!\right]$ with $x_t(S) = 1 < y_t(T)$ to obtain a sequence $x^{\prime}$ such that $x^{\prime}_t\leq y_t(T)$ for all $t\in \N$.
Then, by anonymity we get $x(S) \sim x^{\prime}$, and by asymptotic density-one Pareto $x^{\prime}\prec y(T)$, hence $x(S)\prec y(T)$.

For $x(T)$ and $y(S)$, $t\in [1, t_1!]$, $x_t(T)=y_t(S)=1$,  in view of (\ref{L2E1}), for the coordinates $t\in \left[t_{1}!, t_{2m+3}!\right]\cap\N$, $\left|\left\{t: y_t(S)=1<x_t(T)\right\}\right| < \left|\left\{t: y_t(S)>x_t(T)\geq 1\right\}\right|$, hence, $y_t(S) - x_t(T)\geq 1$ for all $t\in \overset{\infty}{\underset{k=m+2}{\bigcup}} U_k(T)$, and $y_t(S)=x_t(T)=1$ for the remaining $t\in \N$.
Perform a finite permutation of $y(S)$, among coordinates in $\left[1, t_{2m+3}!-\frac{t_{2m+3}!}{t_{2m+2}!}\right]$, and invoke anonymity and asymptotic density-one Pareto to obtain $x(T)\prec y(S)$. 
Hence $x(S)\prec y(T)\prec x(T)\prec y(S)$ or $x(S)\prec y(S)$, therefore $S\in \Gamma$.}


\item{$x(T)\sim y(T)$, i.e., $T\notin\Gamma$:
Drop $t_1$, $t_2$, $t_3$, $t_6$, $t_7$,  $\cdots$, $t_{2m}$ and $t_{2m+1}$ (with $m\geq 3$)  so as to obtain $S :=\{t_4, t_5, t_{2m+2}, t_{2m+3},\cdots\} \in \Omega(T)$ such that $\left|U_1(T)\right| + \left|U_2(T)\right|<\frac{t_7!}{t_6!}+\frac{t_9!}{t_8!}+\cdots+\frac{t_{2m+1}!}{t_{2m}!}$.

For $x(S)$ and $y(T)$, $y_t(T)=x_t(S)= 1$ for all $t\in\left[1, t_2!\right]$; in view of (\ref{L2E1}), for the coordinates $t\in \left(t_{2}!, t_{2m+2}!\right]\cap\N$, we have that $\left|\left\{t: y_t(T)>1\right\}\right| > \left|\left\{t: x_t(S)> 1\right\}\right|$, $y_t(T) - x_t(S)\geq 1$ for all $t\in\overset{\infty}{\underset{k=m+1}{\bigcup}}\left(t_{2k}!, t_{2k+2}!- \frac{t_{2k+2}!}{t_{2k+1}!}\right]\cap\N$, and $y_t(T)=x_t(S)=1$ for the remaining $t\in \N$.
Thus, $x_t(S)\leq y_t(T)$ for all but finitely many $t\in \N$.
A finite permutation, similar to the one described in (b), applied to $x(S)$ yields a sequence $x^{\prime}$ such that $x^{\prime}_t\leq y_t(T)$ for all $t\in \N$.
Then, by anonymity we have that $x(S) \sim x^{\prime}$, $x^{\prime}\prec y(T)$ holds by asymptotic density-one Pareto, hence $x(S)\prec y(T)$.

For $x(T)$ and $y(S)$, for all $t\in [1, t_1!]$, $y_t(S) = 1 = x_t(T)$; in view of (\ref{L2E1}), for the coordinates $t\in \left(t_{1}!, t_{2m+2}!\right]\cap\N$, $\left|\left\{t: y_t(S)>x_t(T)=1\right\}\right| > \left|\left\{t: x_t(T)>y_t(S)> 1\right\}\right|$, $y_t(T) - x_t(S)\geq 1$.
A finite permutation, similar to the one described in (b), applied to $x(T)$ yields a sequence $x^{\prime}$ such that $x^{\prime}_t\leq y_t(S)$ for all $t\in \N$.
Then, by anonymity $x(T) \sim x^{\prime}$, it follows follows from asymptotic density-one Pareto that $x^{\prime}\prec y(S)$, hence $x(T)\prec y(S)$.
Therefore, $x(S)\prec y(T)\sim x(T)\prec y(S)$ or $x(S)\prec y(S)$, thus $S\in \Gamma$.}
\end{enumerate}
\end{proof}


\bibliographystyle{plainnat}
\bibliography{APAnonymity}

\end{document}